\documentclass[hidelinks,conference]{IEEEtran}

\usepackage{graphicx}
   \graphicspath{{./Figures/}}
\usepackage{footnote}   
\usepackage{amssymb}
\usepackage{amsmath}    
\usepackage{algorithm}
\usepackage{algorithmic}
\usepackage{multirow}
\usepackage[numbers,sort&compress]{natbib}
\usepackage{svg}
\usepackage{color}
\DeclareGraphicsExtensions{.pdf,.jpeg,.png,.fig, .emf}
   \usepackage{subfigure}
    \usepackage{epstopdf}
    \usepackage{epsfig}
    \usepackage{subfigure}
    \usepackage{epstopdf}
    \usepackage{epsfig}

\usepackage{RobStd}
\HideNotes
\usepackage{makecell}
\usepackage{booktabs}
\usepackage{multirow}
\usepackage{xspace}

\usepackage{makecell}
\usepackage{siunitx}

\newcommand\eat[1]{}

\begin{document}
\bstctlcite{IEEEexample:BSTcontrol}
\title{TReCiM: Lower Power and Temperature-Resilient Multibit 2FeFET-1T Compute-in-Memory Design}
 \author{
 \small
 Yifei Zhou$^{1}$, Thomas Kämpfe$^{2}$, Kai Ni$^3$, Hussam Amrouch$^4$, Cheng Zhuo$^{1,5,*}$, Xunzhao Yin$^{1,5,*}$\\
 $^1$Zhejiang University, Hangzhou, China;
$^2$Fraunhofer IPMS, Dresden, Germany\\
 $^3$Department of Electrical Engineering, University of Notre Dame, Notre Dame, USA\\
 $^4$Chair of AI Processor Design, Technical University of Munich; TUM School of Computation, Information \\ and Technology; Munich Institute of Robotics and Machine Intelligence, Munich, Germany\\
 $^5$Key Laboratory of Collaborative Sensing and
Autonomous Unmanned Systems of Zhejiang Province, China\\
$^*$Corresponding authors, email: \{czhuo, xzyin1\}@zju.edu.cn
\vspace{-1ex}
}

\maketitle
\pagestyle{empty}

%

\begin{abstract}
Compute-in-memory (CiM) emerges as a promising solution to solve hardware challenges in artificial intelligence (AI) and the Internet of Things (IoT), particularly addressing the "memory wall" issue. By utilizing nonvolatile memory (NVM) devices in a crossbar structure, CiM efficiently accelerates multiply-accumulate (MAC) computations, the crucial operations in neural networks and other AI models. 
Among various NVM devices, Ferroelectric FET (FeFET) is particularly appealing for ultra-low-power CiM arrays due to its CMOS compatibility, voltage-driven write/read mechanisms and high I$_{ON}$/I$_{OFF}$ ratio. Moreover, subthreshold-operated FeFETs, which operate at scaling voltages in the subthreshold region, can further minimize the power consumption of CiM array. However, subthreshold-FeFETs 
are susceptible to temperature drift, resulting in computation accuracy degradation. 
Existing solutions exhibit weak temperature resilience at larger array size and only support 1-bit.
In this paper, we propose TReCiM, an ultra-low-power temperature-resilient multibit 2FeFET-1T CiM design that reliably performs MAC operations in the subthreshold-FeFET region with temperature ranging from 0℃ to 85℃ at scale.
We  benchmark our design using NeuroSim framework in the context of  VGG-8 neural network architecture running the CIFAR-10 dataset. 
Benchmarking results suggest that when considering temperature drift impact, our proposed TReCiM array achieves $91.31\%$ accuracy, with $1.86\%$ accuracy improvement compared to existing 1-bit 2T-1FeFET CiM array.
Furthermore, our proposed design achieves 48.03 TOPS/W energy efficiency at  system level, comparable to existing designs with smaller technology feature sizes.
\end{abstract}

\section{Introduction}
\label{sec:introduction}

The need for processing large amounts of data efficiently has grown rapidly with the advancement of artificial intelligence (AI) and the Internet of Things (IoT). Traditional von Neumann architectures face the "memory wall" challenge, which stems from frequent data transfers and high power consumption between the processing and storage units \cite{Wulf_1995}. 
To address this bottleneck, compute-in-memory (CiM) has emerged as a promising solution for energy-efficient designs \cite{yin2023ultracompact, wei2023imga, yin2024ferroelectric, eldebiky2023correctnet,  shou2023see, yan2022computing, yang2024energy}. CiM enables parallel arithmetic and logical computing directly within the memory, thus eliminating the need for data transfers and enhancing energy efficiency \cite{chen2022accelerating, yin2022ferroelectric, wang2021triangle, huang2023fefet, yan2023improving, Huang2024}.
While numerous CiM designs have been proposed, continuous optimization on energy efficiency is still desirable, particularly given the significant number of multiply-accumulate (MAC) operations required for  neural network processing at the edges.

To further reduce power consumption in CiM designs, two common approaches are employed: (i) The utilization of emerging non-volatile memory (NVM) devices and subthreshold computing. 
NVM devices used in CiM designs include resistive random access memory (ReRAM) \cite{Yu_2021}, phase-change memory (PCM) \cite{Khaddam_2021}, and ferroelectric FET (FeFET) \cite{ni2019ferroelectric, Soliman_2020,cai2022energy, Sk_2023, liu2022cosime, liu2023reconfigurable, xu2023challenges, qian2024cnash, qian2024hycim}. 
FeFETs, in particular, offer advantages such as CMOS compatibility, low leakage current, high I$_{ON}$/I$_{OFF}$ ratio, and energy-efficient voltage-driven write and read operations \cite{Soliman_2020,Sk_2023,li2024febim}. 
(ii) Another approach, subthreshold computing, scales down  the operating voltage of FeFETs to the subthreshold region (subthreshold-FeFET), leading to significant power reductions compared to devices operating in the saturation region.
Combining these two approach results in ultra-low power subthreshold-FeFET based  CiM design.
However, the temperature sensitivity of FeFET devices, especially in the subthreshold region, is often overlooked by designers \cite{amrouch_fefet_device_temperature}. Temperature variations can have a substantial impact on the FeFET device output, resulting in potential circuit errors.

Researches have explored designs of temperature-resilient CiM based on subthreshold-FeFET \cite{zhou2024low}. 
However, existing designs still exhibit weak temperature resilience, and the existing 2T-1FeFET design only supports 1-bit cell, 
limiting the CiM density.
To further reduce power consumption and improve the density while  addressing the temperature drift problem of  subthreshold-FeFET based CiM designs, we propose TReCiM, novel temperature-resilient  2FeFET-1T CiM design.
TReCiM minimizes the impact of temperature variations from 0℃ to 85℃ on the subthreshold-FeFET based CiM array while achieving ultra-low power consumption and supporting multibit CiM computing at scale.
TReCiM employs a 2FeFET clamp  structure and utilizes the complementary-to-absolute-temperature (CTAT) characteristic of MOSFET to achieve the temperature-resilience. 
It enables multibit CiM computing by exploiting the multi-level cell (MLC) characteristics of FeFETs. 
The multibit MAC operation and temperature resilience principles of the proposed design at both the cell and the array levels are demonstrated and validated.
The cycle-to-cycle simulator NeuroSim \cite{NeuroSim} is also modified by embedding the parameters and specifications of  TReCiM at array level to calibrate the CiM behavior at system level.
The proposed design is then benchmarked   
in the context of deep neural network (DNN) accelerators. 
Benchmarking results from NeuroSim shows that the proposed  2FeFET-1T CiM design with 2-bit density achieves $91.31\%$  accuracy for the VGG-8 neural network running CIFAR-10, considering 
the impacts of device-to-device variation. 
The energy efficiency of the CiM system reaches 48.03 TOPS/W.

The remainder of the paper is organized as follows. Section \ref{sec:background} provides a comprehensive review of FeFET and subthreshold computing, states the challenges on temperature-resilience and computing efficiency of emerging designs.
Section \ref{sec:cell} introduces the proposed temperature-resilient and multibit 2FeFET-1T cell and array designs.
Section \ref{sec:Evaluation} validates the design functionality at both the cell and array levels, and further validates the proposed design in the context of VGG neural networks using NeuroSim, benchmarking the design at the system level. Finally, Section \ref{sec:conclusion} concludes.

\section{Background}
In this section, we review the FeFET device and the model capturing the temperature dependency of FeFET, introduce the subthreshold CiM design, and illustrate challenges of temperature-resilience and computing efficiency on existing designs.
\label{sec:background}

\subsection{FeFET basics}
\label{sec:device}
\begin{figure}
    \centering
    \includegraphics[width=1\linewidth]{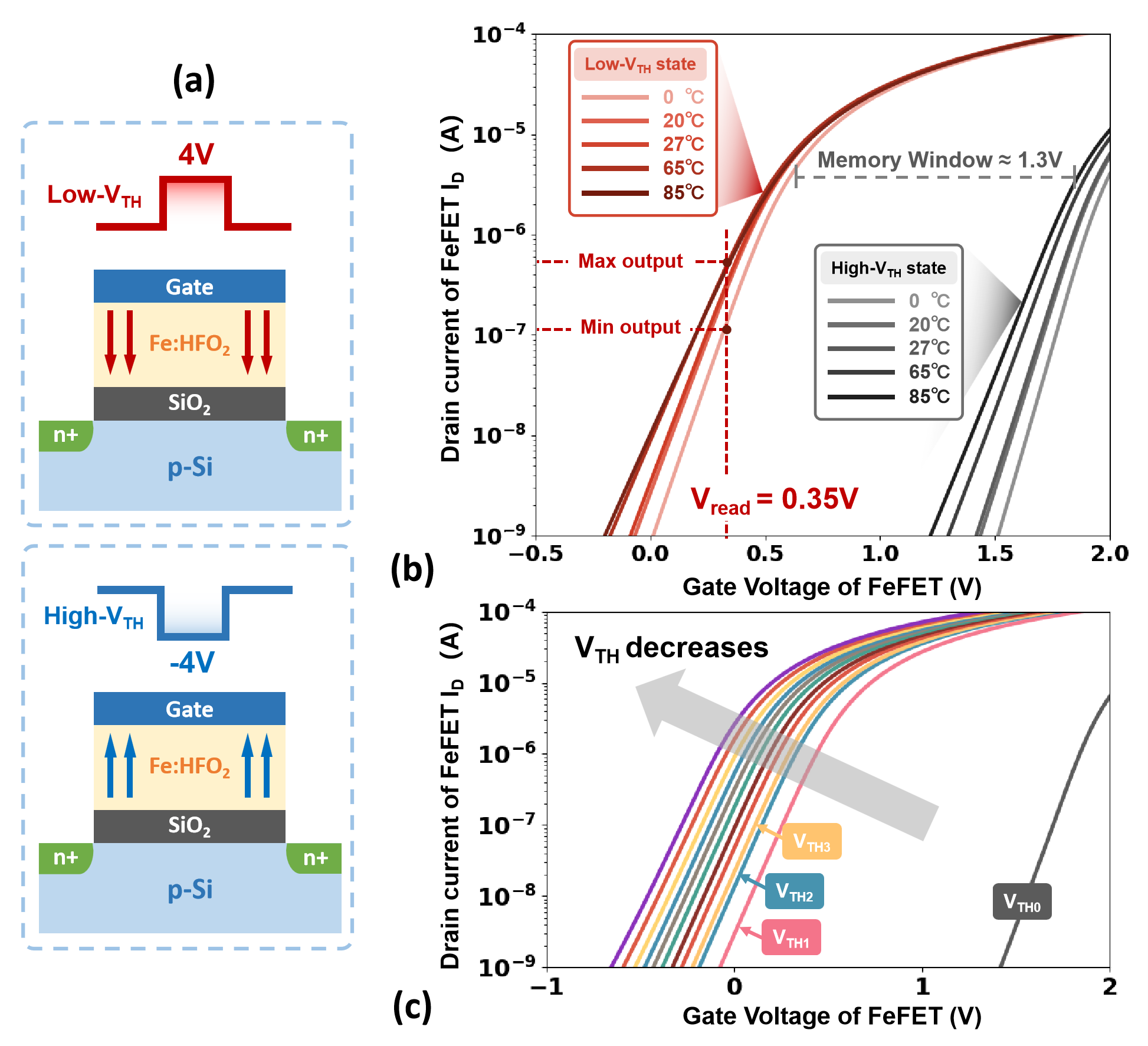}
    \vspace{-2em}
    \caption{(a) FeFET structure and binary states. By applying different  write voltages, the FeFET can be programmed to different polarization directions, resulting in two distinct states. (b) Characteristic  of binary states (low-V$_{TH}$/high-V$_{TH}$) at different temperatures. The variation under operating temperature causes non-negligible output swings. (c) Simulated MLC characteristics with more than 3-bit V$_{TH}$ states based on Preisach FeFET model.}
    \label{fig:fefet}
    \vspace{-2ex}
\end{figure}

FeFETs have been considered as a competitive choice for NVM due to their desirable characteristics such as ultra-low leakage current, high I$_{ON}$/I$_{OFF}$ ratio, voltage-driven mechanisms, and compatibility with CMOS technology \cite{hu2021memory}. FeFETs utilize a ferroelectric film as the gate insulator in a MOSFET, employing $HfO_2$ as the ferroelectric dielectric. Unlike other NVM technologies that require high conduction current for writing, ferroelectric memory relies on the electric field-driven polarization switching process, resulting in lower power consumption during write operations \cite{ni2019ferroelectric}. Fig. \ref{fig:fefet}(a) shows that, by applying positive or negative gate pulses ($\pm$4V) to FeFET devices after activation, ferroelectric polarization switches to point towards the channel or the gate. This attracts electrons or holes in the channel, thus setting the device in a low-V$_{TH}$/high-V$_{TH}$ state \cite{khan2020future}. The stored data ("1" or "0") can be read by measuring the drain current (i.e., I$_{ON}$ or I$_{OFF}$) while applying a lower gate voltage between the low-V$_{TH}$ and high-V$_{TH}$ levels.

Various models have been proposed for FeFET, such as the negative capacitance FET (NCFET) based model \cite{aziz2016physics}, the multi-domain Preisach model \cite{ni2018A}, and the comprehensive Monte Carlo model \cite{deng2020acomprehensive}. However, these models often neglect the effects of temperature on device behavior, limiting their ability to accurately validate the functions and evaluate the performance of FeFET based circuits in practical edge devices. 
Although some studies have explored the impacts of temperature on FeFET devices \cite{amrouch_fefet_device_temperature, thomann2021reliability}, 
the present work goes further by exploring the temperature effects on subthreshold-FeFET based CiM structures \cite{zhou2024low}. 
In this design, we employ the experimentally calibrated Preisach FeFET compact model \cite{ni2018A} in conjunction with the Intel FinFET model. This model considers the ferroelectric film as comprising numerous independent domains, each having its own switching coercive field. The model also takes into account the temperature effects on FeFET devices. 
Fig. \ref{fig:fefet}(b) illustrates the I$_D$-V$_G$ characteristic after a memory write operation with gate voltages of $\pm$4V, showing a memory window of approximately 1.3V.

\subsection{Subthreshold computing}

Scaling the operating voltage to subthreshold region rather than saturation region is a promising way to optimize the energy efficiency of CiM designs.
Subthreshold-FeFETs are acquired by applying scaled read voltage V$_{read}$ between the programmed memory window as shown in Fig. \ref{fig:fefet}(b).
In this paper, we set the read voltage V$_{read}=0.35V$ and in the subthreshold region of the curve corresponding to low-V$_{TH}$ state.
As a result, the FeFET drain current I$_{D}$ is lowered down, achieving significant power reduction. 

However, Fig. \ref{fig:fefet}(b) also demonstrates the significant impact of operating temperatures on the ON current of FeFETs, resulting in considerable deviations between the maximum and minimum CiM outputs. 
While scaling voltages results in reduced power consumption, it also increases the vulnerability of the devices to temperature drift, which can degrade their performance and functionality. Previous designs have made efforts to mitigate temperature impacts on FeFET based ternary content addressable memory (TCAM) designs \cite{thomann2021reliability}. A 2T-1FeFET CiM design presented in \cite{zhou2024low} has demonstrated the possibility of achieving a balance between power consumption and temperature resilience in CiM, though it still exhibits relatively low temperature resilience as the  array size increases. 

\begin{figure*}
    \centering
    \includegraphics[width=1\linewidth]{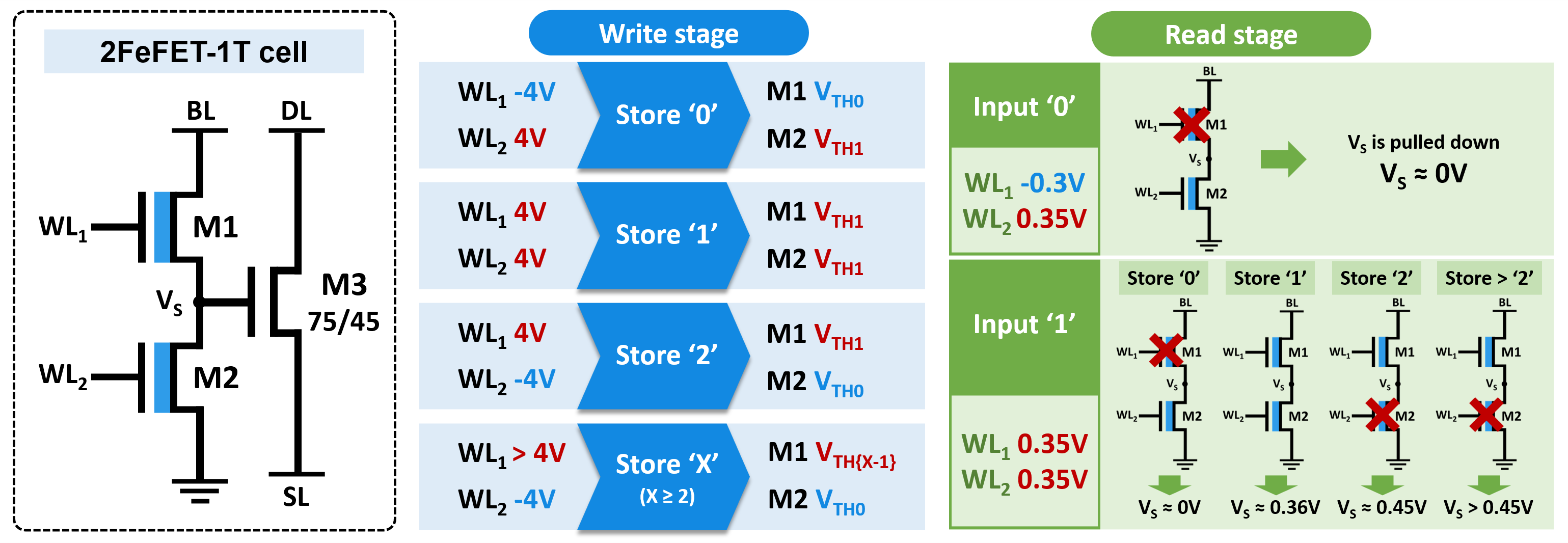}
    \vspace{-2em}
    \caption{Schematic of the proposed 2FeFET-1T cell design for TReCiM array. The cell shown here can store 2-bit data. During 
 the write stages, the stored data is written by voltages on WLs. During the read stages, the 2FeFET-1T cell performs MAC operations with different stored data and inputs.}
    \label{fig:cell}
    \vspace{-1em}
\end{figure*}

 \subsection{Challenges of existing designs}

Work in \cite{zhou2024low} has assessed the significance of temperature-resilient designs in FeFET based CiM architectures, including the 1FeFET-1R structure \cite{Soliman_2020} and the temperature-resilient 2T-1FeFET structure \cite{zhou2024low}. 
Although the 2T-1FeFET design demonstrates a significant reduction in cell output fluctuation compared to the 1FeFET-1R structure, there is still room for further improvement in reducing these fluctuations and supporting larger array sizes.
\cite{zhou2024low} defines the parameters Noise Margin Rate (NMR) and NMR$_{min}$ to evaluate the temperature-resilience of the design, and reports fascinating improvement on NMR$_{min}$ values.
However, the 2T-1FeFET design uses the combination of capacitors as the sense amplifier. The sense amplifier that utilizes capacitors to convert current signals to voltage signals requires additional time for discharging the capacitors after each MAC operation, which leads to higher latencies.
Moreover, as the existing works have explored the application of subthreshold-FeFETs, the MLC characteristics of FeFETs have not been fully exploited to further optimize the computing efficiency.
The MLC characteristics of the simulated FeFET devices with different write pulses are shown in Fig. \ref{fig:fefet}(c), and the MLC V$_{TH}$ characteristics are also shown and labeled.
These challenges call for potential improvement in the design of CiM.

\section{TReCiM design
}
\label{sec:cell}

In this section, we present our approach to enhance the robustness and density of low power CiM arrays by introducing  TReCiM, the temperature-resilient and multibit 2FeFET-1T design.
We conduct the discussions on our proposed design at both  cell and array levels, highlighting its advantages over existing designs. 
The evaluations are performed considering various cell precisions, taking into account the temperature drift ranging from 0℃ to 85℃ and device  variation. 

\subsection{Proposed 2FeFET-1T cell}
\label{propose}

Fig. \ref{fig:cell} shows the proposed  multibit 2FeFET-1T TreCiM cell design.
The structure consists of two  FeFETs connected in  a clamp  structure to control the intermediate voltage at node V$_S$, which also serves as the gate voltage of the NMOS M3.
The two gates of FeFETs are controlled separately by WL$_1$ and WL$_2$,  serving as the inputs to the cell. The source  of MOSFET works as the output and connects to SL.
We define the weighted states (the states stored in the cell) during the write stage as follows:

1.	If  M2 is in the V$_{TH1}$ state (labeled in Fig. \ref{fig:fefet}(c)), while M1 is in the V$_{TH0}$ state (the high threshold voltage in Fig. \ref{fig:fefet}(c)), i.e., WL$_1=-4V$ and WL$_2=4V$ during the write. This state is denoted as '0'.

2.	If both FeFETs are in the V$_{TH1}$ states, i.e., WL$_1=4V$ and WL$_2=4V$ during the write, the intermediate output V$_S$ of  FeFETs is solely determined by the inputs. 
This state is denoted as '1'.

3.	If M1 is in the V$_{TH1}$ state, while M2 is in the V$_{TH0}$ state, i.e., WL$_1=4V$ and WL$_2=-4V$ during the write. 
This state is denoted as '2'.

4. More states can be stored with the similar approach as state '2'. To store state 'X', we set M1 to the V$_{TH\{X-1\}}$ state, and M2 to the V$_{TH0}$ state. For example, state '3' is set by WL$_1=4.13V$ and WL$_2=-4V$ during the write.


During the read or computation stages, the 2FeFET-1T cell performs MAC operations with binary inputs and weighted data stored in the cells. 
Fig. \ref{fig:cell} shows the different intermediate voltages at V$_S$ corresponding to different inputs. 
The voltage at V$_S$ is primarily determined by M1 and M2. During the read stages, the DL is set to 0.4V, while the BL is set to 1V.
When the input is '0', we apply -0.3V to the gate of M1 to turn off M1, and  0.35V to the gate of M2 to enable M2, 
pulling V$_S$ to ground. Consequently, the output of the cell will be close to 0 regardless of the stored data. 
On the contrary, when the input is '1', we apply  0.35V to both M1 and M2 gates.
The MAC results are dependent on the stored states as follows:

1. If '0' is stored, only M2 is enabled. Thus, the voltage at V$_S$ is pulled down to ground, resulting in a small output on the SL. 

2. If '1' is stored, M1 and M2 are both enabled, working as a voltage divider. The voltage at V$_S$ is stably clamped by both FeFETs, ensuring its resilience to temperature variations. 

3. Furthermore, when the stored state is equal to or greater than '2', M1 is enabled while M2 is turned OFF.
The different stored values  correspond to the MLC V$_{TH}$ states of M1, resulting in distinct levels of drain currents given the same input voltage. 
As a result, the voltage at V$_S$ changes, leading to different  currents flowing through  M3 to  SL.

\begin{figure}
    \centering
    \includegraphics[width=1\linewidth]{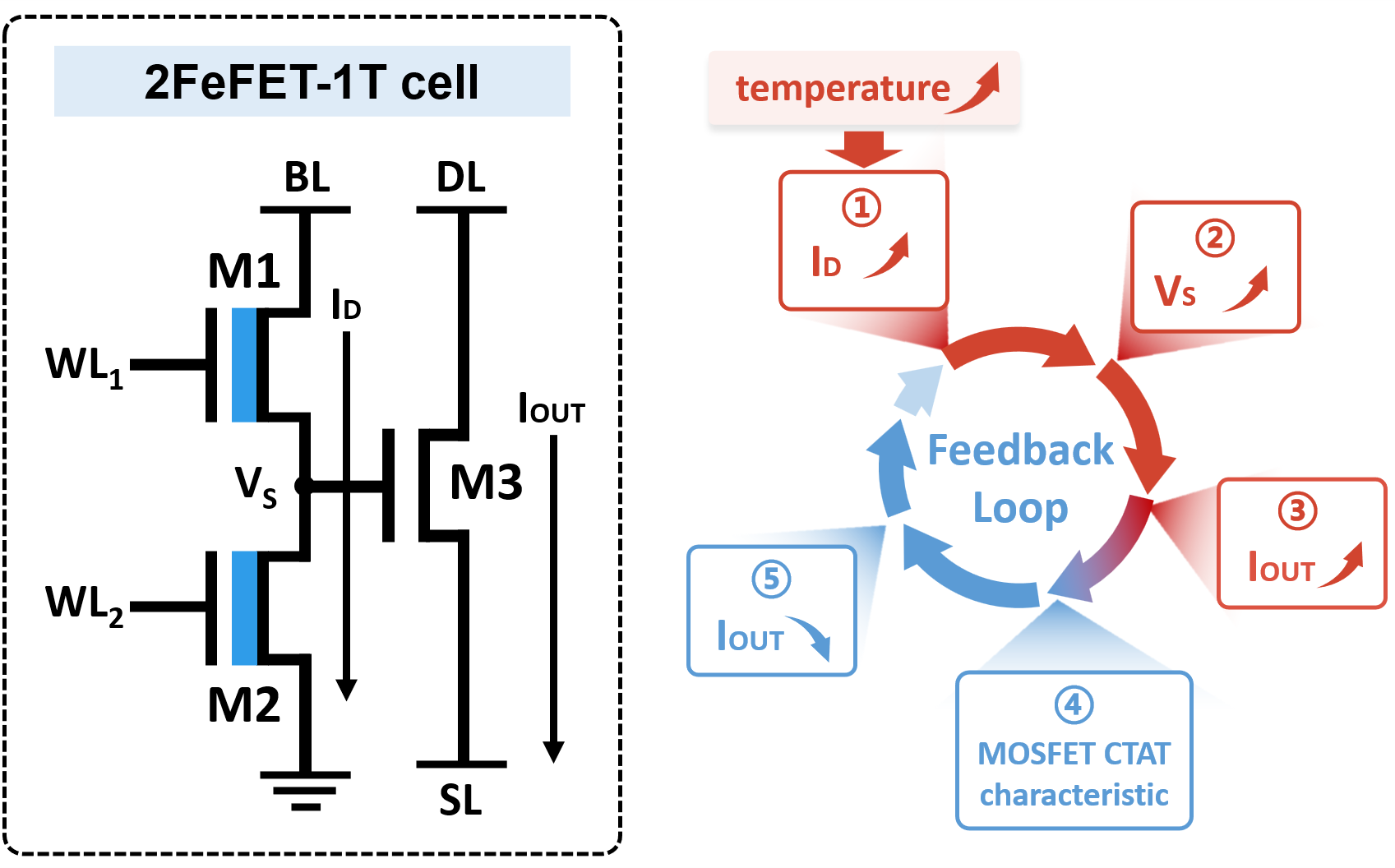}
    \vspace{-4ex}
    \caption{The feedback loop for temperature drift compensation. The output current drop will also rise back as the temperature decreases.}
    \vspace{-2ex}
    \label{fig:loop}
\end{figure}

The states greater than '1' create a non-clamped equivalent circuit, which is susceptible to temperature drift. 
To enhance the temperature resilience of the cell, parameters such as the W/L (width/length) ratio, read latencies, and write latencies are adjusted to optimize the cell. 
Moreover, the MOSFET in the proposed 2FeFET-1T cell structure exhibit a CTAT characteristic, contributing to its temperature resilience. 
The CTAT characteristic of the MOSFET can be expressed by the  drain current formula in subthreshold region
$$I_D=I_0 exp(\frac{V_{gs}}{\xi V_{T}})$$
with $V_T=\frac{kT}{q}$, and $\xi$ is a non-ideal constant relevant to device technology. 
The drain current of MOSFET is complementary to the absolute temperature.
In this structure, as shown in Fig. \ref{fig:loop},  a feedback mechanism is employed to reduce the impacts caused by the temperature drift.
As the temperature increases, the drain current of M1 also increases. 
Due to the reason that M2 is disabled and equivalent to a large resistor, the voltage at V$_S$ rises, causing an increase in the output current of M3. 
However, due to the CTAT characteristic, the extent of this current increase is reduced, resulting in smaller output fluctuations. 
Conversely, as the temperature decreases, the output current drop will also rise back partially, maintaining the stability and the accuracy of the computation.

\begin{figure}
    \centering
    \includegraphics[width=0.8\linewidth]{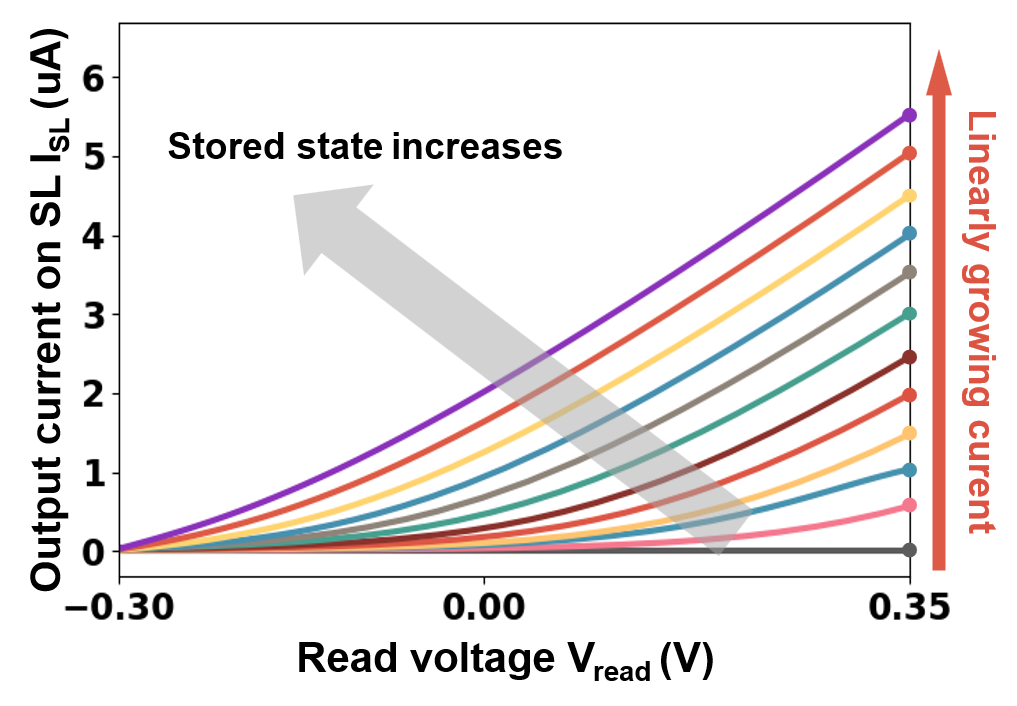}
    \vspace{-1em}
    \caption{Simulated I$_{SL}$-V$_{read}$ characteristics of the 2FeFET-1T cell with varying states stored in the cell. The cells with different stored states show linearly growing current output at given operating voltage.}
    \label{fig:cell_cl}
\end{figure}

We hereby focus  on the design and validation of the proposed 2FeFET-1T design storing 2 bits. 
Note that similar approaches can be employed to store and compute more weighted bits by programming MLC states as shown in Fig. \ref{fig:fefet}(c)  with larger write voltages on WL$_1$. 
Fig. \ref{fig:cell_cl} depicts the I$_{SL}$-V$_{read}$ characteristics of the proposed 2FeFET-1T cell with varying weighted states. The curves represent different stored states within the cell,
and the current results demonstrate that by adjusting the threshold voltage V$_{TH}$ states of the FeFET devices in the 2FeFET-1T cell,
the cell is capable of  storing multibit states and generating corresponding intermediate current outputs.

\subsection{Proposed 2FeFET-1T CiM array}

\begin{figure}
    \vspace{-2ex}
    \centering
    \includegraphics[width=0.75\linewidth]{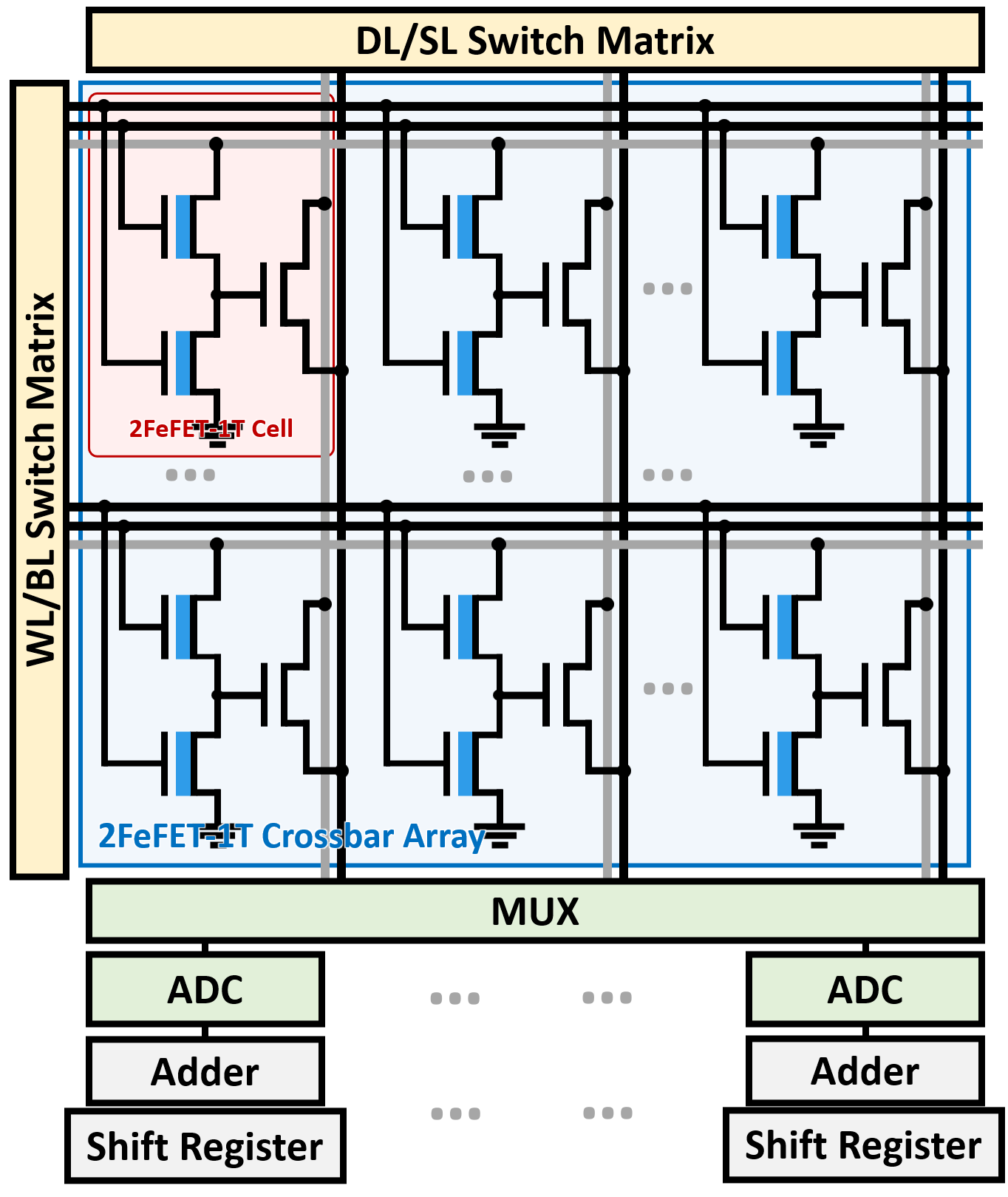}
    \vspace{-0.5em}
    \caption{Macro organization of 2FeFET-1T CiM system. The  CiM system consists of CiM array, ADC and peripheral circuits (MUX, adder, etc.).}
    \label{fig:system}
    \vspace{-2ex}
\end{figure}

Fig. \ref{fig:system} shows the overall schematic of the 2FeFET-1T TReCiM based system. 
We build the subthreshold-FeFET based CiM array with 8 rows. 
Within this array design, we employ a flash analog-to-digital converter (ADC) utilizing current sense amplifiers (CSAs) to minimize the sensing latency \cite{neuro1.4, Schinkel2007}. 

During the write operation of the proposed 2FeFET-1T CiM array, a -4V pulse 
is applied to the WLs to set cells to state '0'. 
Subsequently,  positive pulses corresponding to multibit threshold voltage states
are applied to set the FeFETs to other states, and the magnitudes of the   pulses are determined by the data intended to be stored in the cells. 
During the read/MAC operation, the DLs of the cells are set to 0.4V, while the BLs are set to 1V. 
The  cells within the same row are controlled by two input WLs, and the SLs of cells within the same column are connected to the shared ADC via a MUX. 
When the  input is '0', WL$_1$ is set to -0.3V, and WL$_2$ is set to 0.35V. 
When the input is '1', both WLs are set to 0.35V.
Each column performs the analog parallel read-out operation for MAC computation  between the input voltage and the stored states of cells along the column.
The results of the MAC operations are represented as aggregate currents on the SLs, and are  fed into the ADC for further processing.
Based on the  subthreshold-FeFET computing scheme, the 2FeFET-1T array design also shows ultra-low energy consumption. 
In this array design, we utilize 3-bit flash ADC. 
The flash ADC incorporates a minimum of 7 comparators to achieve a 3-bit precision. 
When analyzing the energy consumption across four different stored states, it is observed that the average energy required for one read operation in the CiM array column and corresponding ADC is approximately 1.36 pJ. The energy consumption of the array is primarily influenced by the scale of the ADC, while the number of CiM array rows has only a minor impact on the overall energy consumption.


\section{Validations and Evaluations}
\label{sec:Evaluation}
In this section, we validate and evaluate our proposed TReCiM design regarding the functionality, robustness, and performance.
We then benchmark the TReCiM array using NeuroSim, a system level simulator for efficiently simulating and validating the CiM architectures. 
We revise the NeuroSim to support temperature resilient subthreshold-FeFET based CiM design, and benchmark the TReCiM 
in the context of VGG-8 neural network architecture running the CIFAR-10 dataset, and compare it with other designs.



\begin{figure}
    \centering
    \includegraphics[width=1\linewidth]{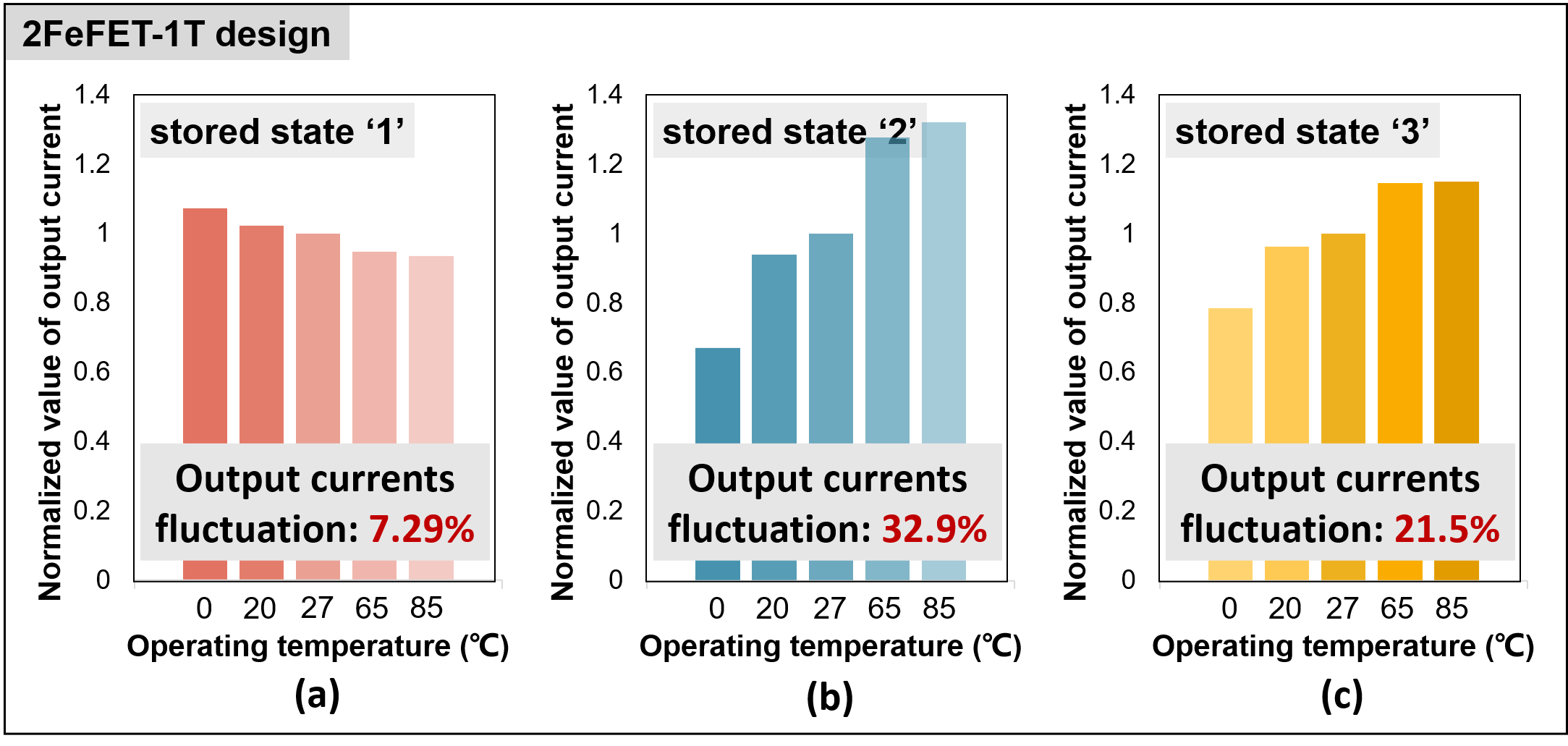}
    \vspace{-2em}
    \caption{Normalized value of 2-bit 2FeFET-1T cell output currents at different temperatures for different stored values. The reference temperature is set at 27℃ (RT).}
    \vspace{-1em}
    \label{fig:eval_cell}
\end{figure}

\subsection{Temperature resilience validation}
We first analyze the temperature resilience of TReCiM cell. 
We assess the impact of temperature drift on the CiM design based on subthreshold-FeFET. 
Fig. \ref{fig:eval_cell}(a)-(c) show the cell output fluctuations under varying temperatures for different stored values '1', '2', and '3', respectively. 
The experiments are conducted with identical input voltages, and the reference temperature is set at 27℃ (RT). 
From the results, 
significant reduction in temperature drift effects during the MAC operations can be observed for the stored value  '1'.
However, for other stored states, output fluctuations remain non-negligible, with the largest fluctuation observed for state '2' at $32.9\%$. 
As the stored values further increase, the V$_{TH}$ value of the FeFET device decreases, approaching the operating point to  the saturation region given the same operating voltage. 
Consequently, the sensitivity of FeFET M1 in Fig. \ref{fig:cell} to temperature drift decreases. 
Therefore, if the cell is used to store binary states, the impact of temperature drift  can be considered negligible.
For the cell storing beyond binary (2-bit or more bits), the sensitivity to the temperature drift is  weakened, and comparable to  other designs which store binary states. 
For example, the current fluctuation over the reference temperature for the binary 2T-1FeFET cell is $26.6\%$, while the current fluctuation for the subthreshold 1-bit 1FeFET-1R cell reaches $52.1\%$.

\begin{figure}
    \centering
    \includegraphics[width=1\linewidth]{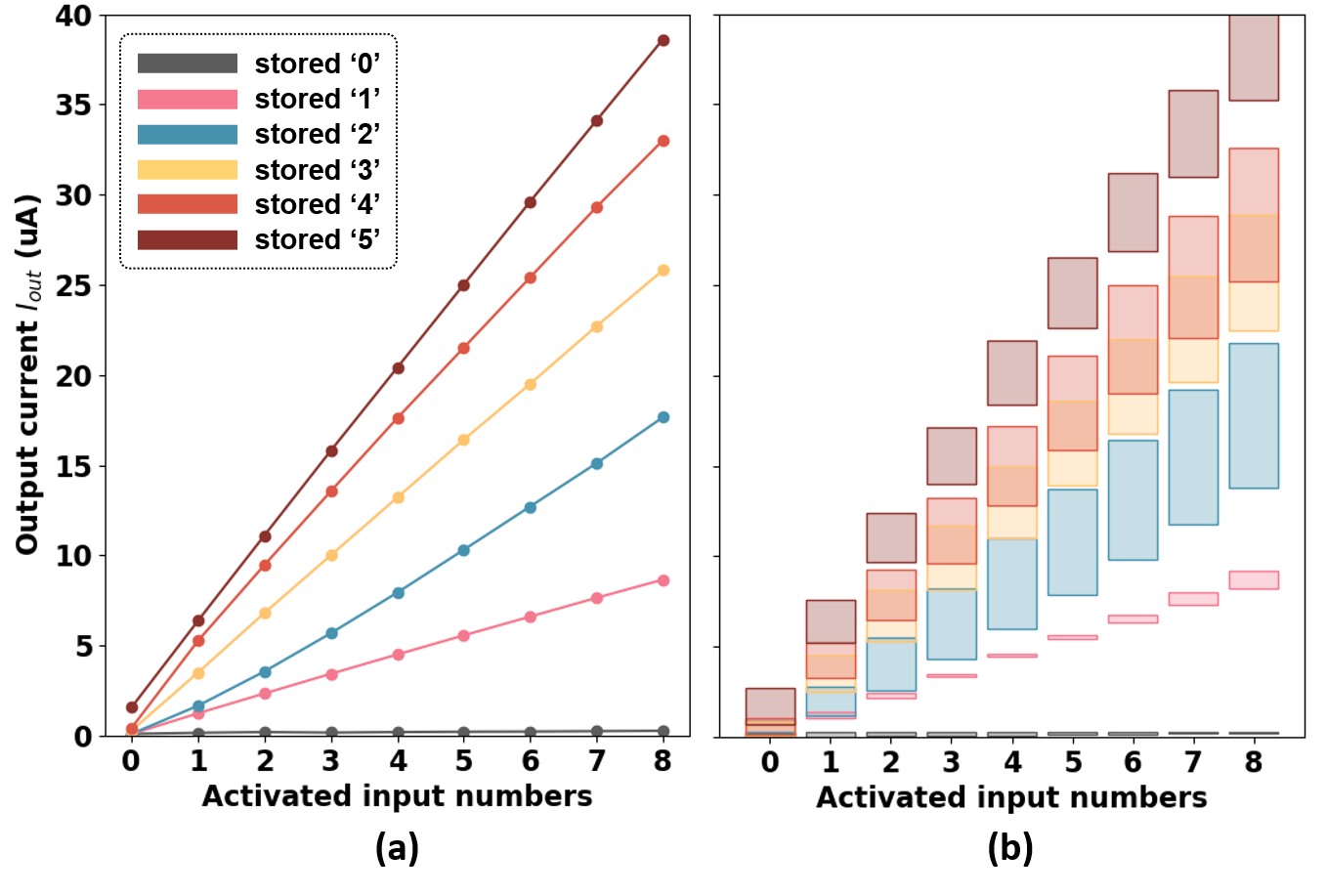}
    \vspace{-5ex}
    \caption{All possible MAC outputs of the proposed 2FeFET-1T TReCiM array containing 8 cells per column, corresponding to the activated input numbers. (a) shows the current outputs at 27°C, which has high linearity. (b) shows the MAC output range at different stored states, with temperature range of 0-85℃.}
    \vspace{-1.5em}
    \label{fig:eval_array}
\end{figure}

Then we validate the temperature resilience of TReCiM array. 
The 2-bit  TReCiM array with 8 cells per column is validated. 
Without loss of generality, all cells within the array are assumed to store the same state.
Fig. \ref{fig:eval_array}(a) illustrates the linearity of MAC outputs across stored states ranging from '0' to '5', and shows all possible MAC results. The experimental measurements are conducted at a temperature of 27℃. In the experiments, a specific input is activated means that we set the corresponding input value to '1'.
Fig. \ref{fig:eval_array}(b) further presents the range of MAC outputs for our proposed TReCiM array across a temperature range of 0-85℃ with different stored states. 
For the 2-bit TReCiM array, we only use the first four stored states.
The output distributions corresponding to the four stored  states are  illustrated. 
The results  show that our proposed design produces distinguishable MAC outputs.
While the overlaps between different MAC outputs cannot be completely eliminated when larger weights are stored, comparing to emerging designs, the extent of the overlap has been significantly decreased due to the application of the clamp  structure and CTAT characteristic of MOSFET.
Fig. \ref{fig:eval_array}(b) also validates that the sensitivity to the temperature drift of TReCiM above state '2' is weakened, showing improvement on output overlaps.
The  results depicted in Fig. \ref{fig:eval_array} demonstrate that the binary 2FeFET-1T TReCiM array  achieves a NMR$_{min}$ value of NMR$_{min}=$NMR$_{7}=0.29$\footnote{larger NMR$_{min}$ value indicates better temperature-resilience.}.
If we consider only the temperature range of 20-85℃, the NMR$_{min}$ increases to NMR$_{min}=$NMR$_{7}=2.6$. 
As for the 2-bit 2FeFET-1T array, the NMR$_{min}= -0.73$. 
Here we use the ratio of 
$$G_{21}=\frac{NMR_{min2}+1}{NMR_{min1}+1}$$ 
to quantify the  improvement of the temperature resilience. 
The 1-bit TReCiM design shows  $3\times$ improvement over the subthreshold 1FeFET-1R design 
from 0 to 85°C \cite{Soliman_2020}, and  $1.06\times$ improvement over the 2T-1FeFET design \cite{zhou2024low}, respectively.
These results validate the effectiveness of the proposed TReCiM in enhancing the temperature resilience.

\begin{figure}
    \centering
    \vspace{-0.5em}
    \includegraphics[width=1\linewidth]{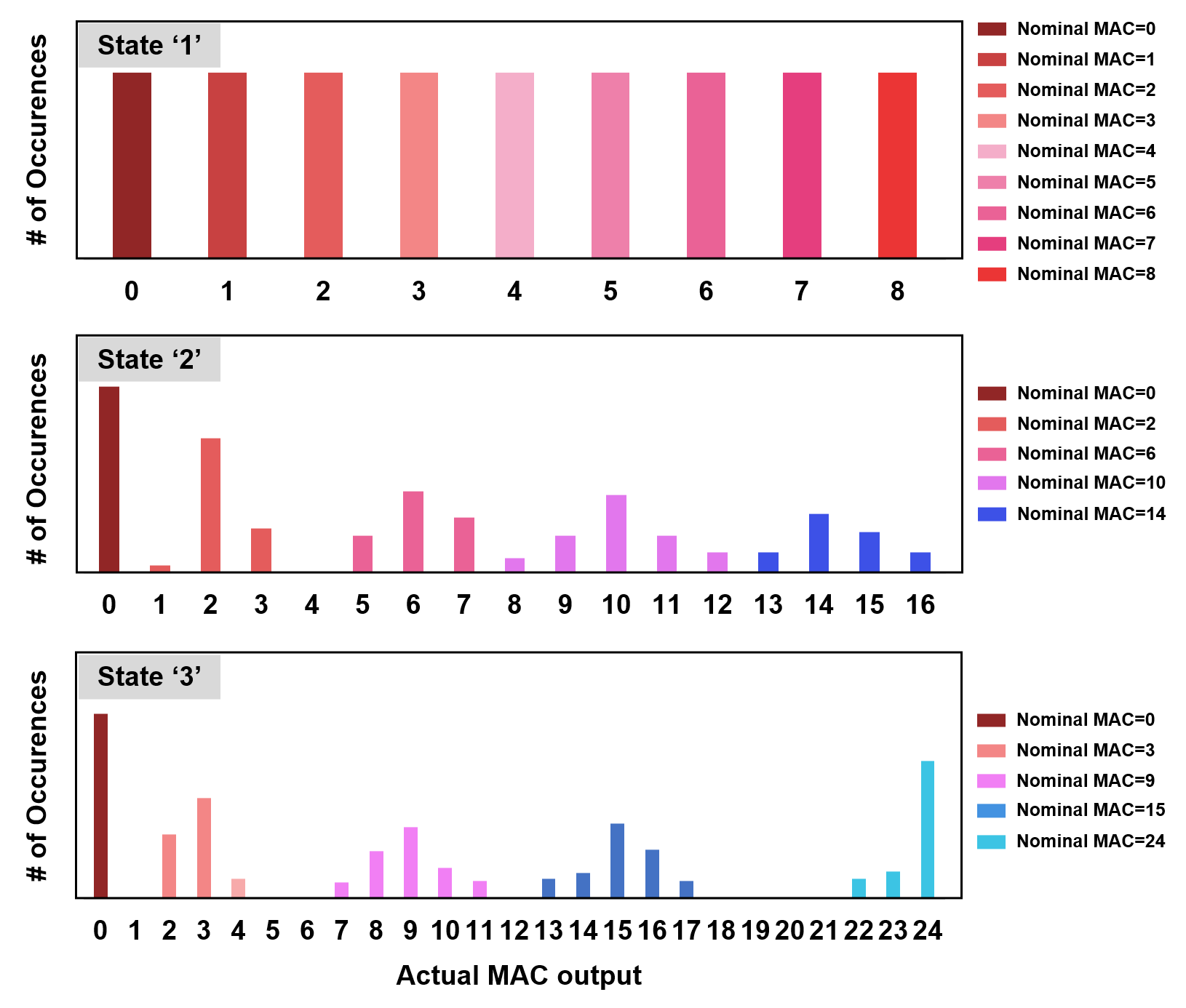}
    \vspace{-2.5em}
    \caption{The impact of process variation with $\sigma_{V_T}=54mV$ at 27°C on 2-bit 2FeFET-1T TReCiM  array digital output via 500 Monte Carlo runs for three  stored states.}
    \vspace{-2ex}
    \label{fig:vari}
\end{figure}

To comprehensively evaluate the impact of process variations on the TReCiM design, we conduct 500 Monte Carlo simulations on the TReCiM array  consisting of the 2-bit TReCiM array and the associated ADC. 
The simulations incorporate an experimental FeFET Gaussian variability of $\sigma_{V_T}=54mV$ and are performed at a temperature of 27°C. 
The output of MAC results are generated  by ADC.
Fig. \ref{fig:vari} shows the MAC output results for  each stored states, revealing that the error caused by the process variations at array level remains in a reasonable range. 
The processes of MAC operation and ADC conversion for state '1' show $100\%$ accuracy.
Cell variations caused by process variation are extracted, with the reference temperature set at 27°C.
In this design, different stored states show varied impacts caused by the process variation. From results in Fig. \ref{fig:vari}, the variation for states '1', '2' and '3' are $3.89\%$, $20.8\%$ and $17.1\%$, respectively.

\subsection{Benchmarking setup}


\begin{table}[t!]
\centering
\caption{NeuroSim simulation settings}
 \vspace{-3mm}
\label{tab:macro_set}
\centering
    \includegraphics[width=1\linewidth]{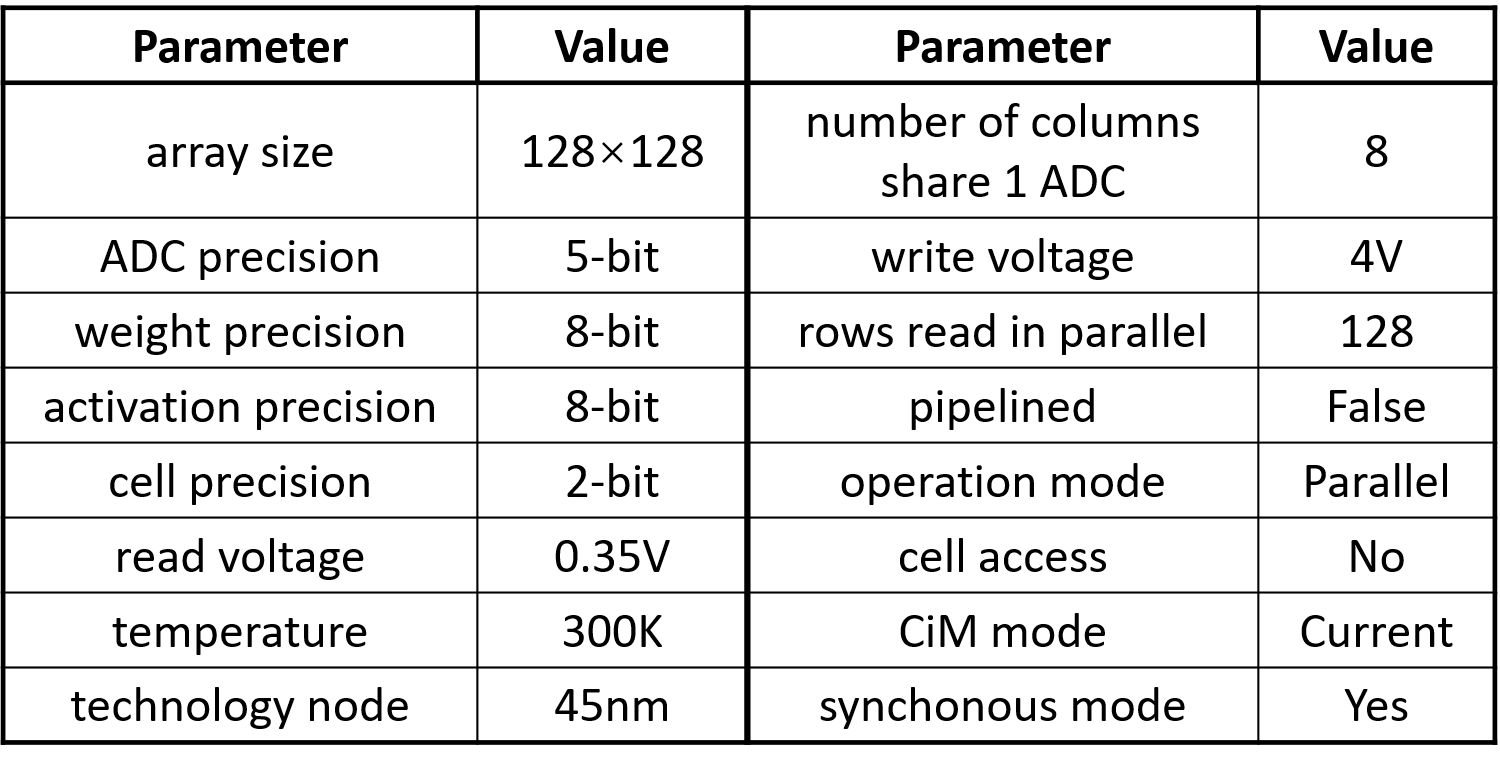}
    \vspace{-3em}
\end{table} 

\begin{table}[t!]
\centering
\caption{Validation results of modules in the sub-array}
 \vspace{-3mm}
\label{tab:peripheral_bench}
\centering
    \includegraphics[width=1\linewidth]{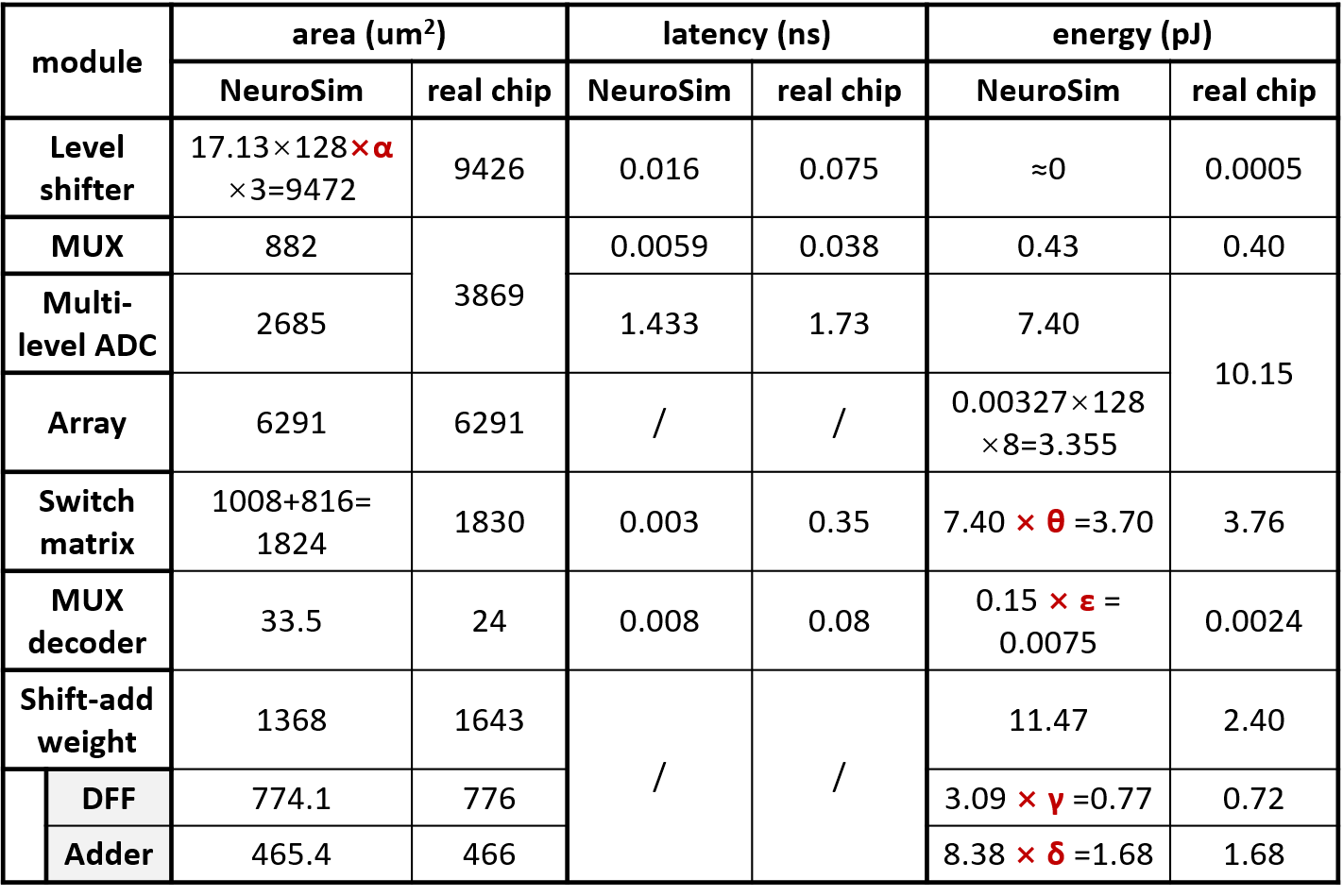}
    \vspace{-3em}
\end{table}

We benchmark our subthreshold-FeFET TReCiM design based on  NeuroSim, a CiM architectural simulator.
NeuroSim is a comprehensive simulation platform specifically designed for efficient and accurate simulation of CiM systems \cite{DNNNeuro}. It employs a hierarchical structure that encompasses various levels of the system. At the lower levels, NeuroSim focuses on simulating the behavior and characteristics of memory cells and transistor technologies. These components are then integrated to form sub-arrays. The next level involves combining sub-arrays and peripheral circuits to create functional units called tiles, which contain multiple processing elements (PEs). The interconnect routing within a tile can utilize either X-Y bus interconnect or H-tree based routing. Finally, multiple PEs are combined to form a complete chip in NeuroSim \cite{neuro1.4}. By processing the simulation procedure from the bottom to the top, NeuroSim accurately captures the behavior and performance of CiM systems as they would manifest in physical implementations. 

The MOSFET transistor models are adopted using the Predictive Technology Model (PTM),  a widely used industry-standard model. 
In this validation, we 
extract the key transistor parameters of the 2FeFET-1T cell from SPICE simulation results.
For the 2FeFET-1T cell, the extracted on/off ratio is 238, on resistance is 118k$\Omega$, and V$_{read}=0.35V$.
To support subreshold-FeFET computing scheme, lower source voltages are employed. Throughout the entire validation process, we maintain V$_{dd}=0.8V$ for the proposed 2FeFET-1T, and subthreshold 2T-1FeFET and 1FeFET-1R CiM designs. For the saturated 1FeFET-1R CiM design, V$_{dd}=1.0V$ is retained.

Fig. \ref{fig:system} provides an overview of the system's structure.
In this CiM structure, considering that the area overhead of ADC is much larger than a column of CiM array, multiple array columns (8 columns in this case) share a single ADC.
During the inference process, NeuroSim normalizes the weights and inputs to integers with a specified precision. In this design, a 2FeFET-1T cell stores 1-bit or 2-bit data, and 8-bit weights are formed by combining 8 or 4 cells. If the input is set to 8-bit precision, it is divided into 8 consecutive "0-1" pulses. To efficiently handle input sparsity, an input sparsity-aware controller counts the number of '1's in the input vector. Once the counter reaches a predefined threshold, the corresponding rows are asserted in parallel \cite{neuro1.4}. 
The default simulator settings for the CiM macros input to NeuroSim are summarized in Table \ref{tab:macro_set}.

Table \ref{tab:peripheral_bench} presents the validation results of the chip's sub-array modules, obtained from NeuroSim before calibration and Cadence SPICE. The results should be aligned for the novel supporting structure in NeuroSim. To ensure accurate area estimation, a wiring area factor $\alpha$ is introduced. For the level shifter, $\alpha$ is set to 1.44 to account for additional wiring and interconnects in the layout \cite{neuro1.4}.
In terms of energy consumption of CiM arrays, the energy consumption results in the table only represent the outcomes for one read. When considering the activation of all 128 rows and 8 columns in one complete operation, NeuroSim predicts an energy consumption of 3.355pJ for CiM array and 7.40pJ for the ADC, which are close to real situations.
Activity factors are also calibrated for DFFs, the adder, the switch matrix, and the control circuits. For DFFs, $\gamma$ is set to $25\%$ to simulate actual switching activity of DFFs, while $\delta$ is set to $20\%$ for the switching activity of adders \cite{neuro1.4}. The switch matrix adopts an activity factor of $\theta=0.5$, and the control circuits use a factor of $\epsilon=5\%$ to account for their switching activity accurately during simulation.

\subsection{Benchmarking results}
\begin{figure*}
\centering
    \includegraphics[width=1\linewidth]{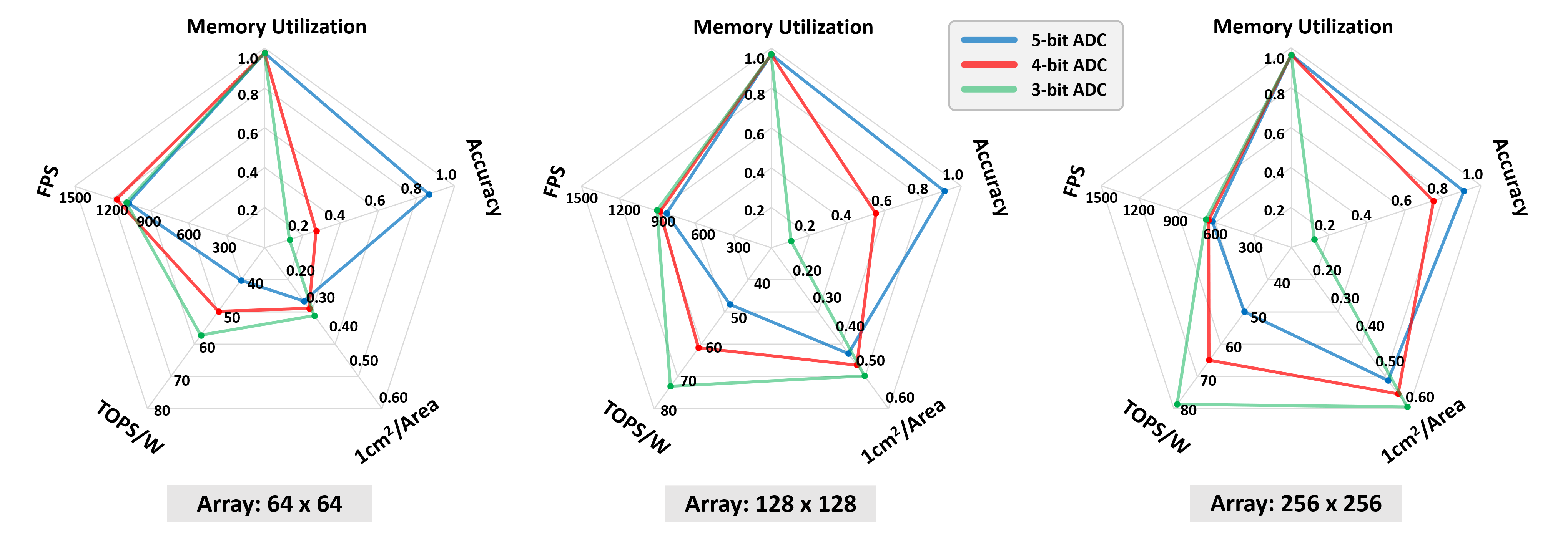}
    \vspace{-5ex}
    \caption{Comparison of inference accuracy, memory utilization, area, energy efficiency and throughput, across different CiM array sizes with 2-bit cell storage precision for VGG-8 running CIFAR-10 dataset.}
    \label{fig:radar}
    \vspace{-1em}
\end{figure*}

\begin{table}[t!]
\centering
\caption{The structure of VGG-8 executed on CIFAR-10 dataset}
 \vspace{-3mm}
\label{tab:vgg8}
\resizebox{1\columnwidth}{!}{
\begin{tabular}{|c|c|c|c|}
\hline
Layer& Input Map & Output Map & Non Linearity \\ \hline
128 3$\times$ 3 Conv1& 32$\times$32$\times$3   & 32$\times$32$\times$64 & ReLU \\ 
\hline
128 3$\times$ 3 Conv2& 32$\times$32$\times$128   & 32$\times$32$\times$128 & ReLU \\ 
\hline
$[2,2]$ MaxPool1 & 32$\times$32$\times$128   & 16$\times$16$\times$128& $-$  \\
\hline
256 3$\times$ 3 Conv3& 16$\times$16$\times$128   & 16$\times$16$\times$256 & ReLU \\ 
\hline
256 3$\times$ 3 Conv4& 16$\times$16$\times$256   & 16$\times$16$\times$256 & ReLU \\ 
\hline
$[2,2]$ MaxPool2 & 16$\times$16$\times$256 & 8$\times$8$\times$256& $-$  \\ 
\hline
512 3$\times$ 3 Conv5& 8$\times$8$\times$256   & 8$\times$8$\times$512 & ReLU \\ 
\hline
512 3$\times$ 3 Conv6& 8$\times$8$\times$512   & 8$\times$8$\times$512 & ReLU \\ 
\hline
$[2,2]$ MaxPool3 & 8$\times$8$\times$512 & 4$\times$4$\times$512 & $-$  \\ 
\hline
8192$\times$8192 FC1 & 1$\times$1$\times$8192 & 1$\times$1$\times$8192 & ReLU  \\ 
\hline
8192$\times$10 FC2 & 1$\times$1$\times$8192 & 1$\times$1$\times$10 & $-$  \\ 
\hline
\end{tabular}
}
\vspace{-4mm}
\end{table}

\begin{table*}
\centering
\vspace{-0.5em}
\caption{Benchmarking results of CiM systems on VGG-8 for CIFAR-10}
 \vspace{-3mm}
\label{tab:benchmark}
\centering
    \includegraphics[width=1\linewidth]{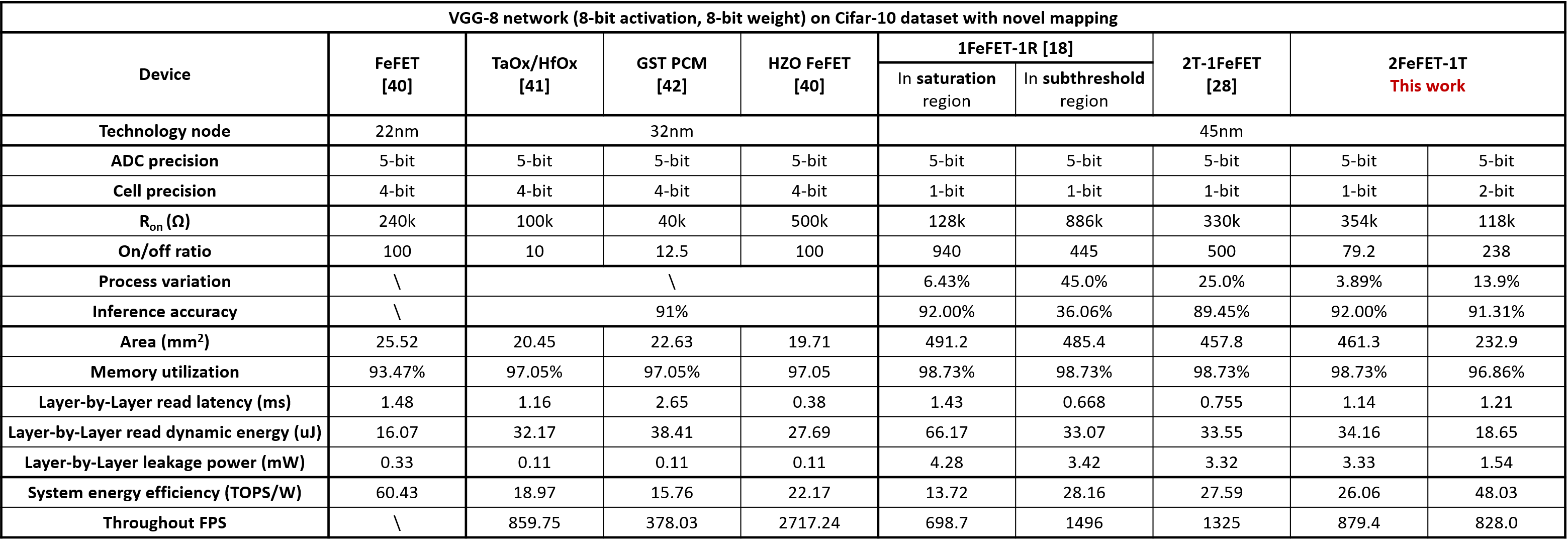}
    \vspace{-3em}
\end{table*}
Here  we benchmark the performance of the proposed TReCiM system using the revised NeuroSim and compare it with other existing designs.
To accommodate the subthreshold computing scheme, scaled  voltages are employed. 
We set V$_{dd}=0.8V$ for subthreshold designs, while the saturated 1FeFET-1R CiM design retains V$_{dd}=1.0V$.
We first evaluate the accuracy of TReCiM in the context of the VGG-8 network architecture running the CIFAR-10 dataset.
The specific VGG-8 network used in the benchmarking is summarized in Table \ref{tab:vgg8}. 
In this study, we adopted the algorithm proposed in DNN+NeuroSim \cite{DNNNeuro}, which assumes that the neural network model or weights are pre-trained off-chip and then mapped onto the TReCiM based inference chip. 
The DNN model is quantized using the WAGE model \cite{wu2018training} to simulate the hardware computation process. Furthermore, the algorithm incorporates simulation capabilities that consider the computation variance of the TReCiM cells. 
For this experiment, we utilized the simulation setups including array sizes, ADC, weight and activation precision listed in Table \ref{tab:macro_set}.
We also consider the  impacts of process variation.
The output fluctuation 
is considered as the TReCiM cell variance. 
For the 2-bit TReCiM design, different stored states lead to different fluctuations in output currents.
To estimate the combined variance for the TReCiM  design, we count the number of each weighted states in the inference for the VGG-8 network model trained on the CIFAR-10 dataset.
The total percentages of weights '0', '1', '2' and '3' in the inference process are $27.2\%$, $24.1\%$, $23.5\%$ and $25.2\%$, respectively. 
To reach a more realistic estimation, we only estimate by variance of weights '1', '2' and '3'. The reason is that, 
The conduction of state '0' is much lower than other states, thus the variation on state '0' has negligible impacts on the inference accuracy. 
Therefore, 
given the percentage of each weights, the average combined  variance for the 2-bit TReCiM  array amounts to $13.9\%$, 
and the inference accuracy reaches $91.31\%$. 
When employing TReCiM cell with binary storage, the output variance reduces to $3.89\%$ and the inference accuracy reaches $92.00\%$, which is comparable to the accuracy without considering the device variability.
Therefore, the ability of TReCiM to mitigate the impacts of
process variation shows a significant advantage, making the proposed CiM a robust and reliable solution for neural network applications.

Fig. \ref{fig:radar} illustrates the performance metrics of various CiM architecture with varying array size and the ADC bit  precision. 
The testing results were based on a layer-by-layer process. 
Increasing the array size leads to a smaller chip area but worse throughput due to large column currents and associated parasitic  load capacitance. 
Conversely, decreasing the ADC precision results in higher energy efficiency due to less required current comparators, but it also leads to accuracy degradation. 
The radar plot demonstrates that the $128\times 128$ and $256\times 256$ array  with a 5-bit ADC achieves a relatively balanced trade-off among accuracy, energy efficiency, throughput, area, and memory utilization.

Table \ref{tab:benchmark} presents the benchmarking results across different devices obtained from NeuroSim. 
All the devices operate in parallel read-out mode. 
From the benchmarking results, our proposed TReCiM design is evaluated using the configuration summarized in Table \ref{tab:macro_set}. The proposed 2FeFET-1T, prior 2T-1FeFET and subthreshold 1FeFET-1R design uses V$_{read}=0.35V$, 
while the saturated 1FeFET-1R design uses V$_{read}=1.3V$. The comparisons show that, the subthreshold 1FeFET-1R design experiences severer impacts caused by process variation, leading to significantly lower inference accuracy.
The 2T-1FeFET design only supports 1-bit cell precision, whereas our proposed 2FeFET-1T CiM design maintains high inference accuracy both at 1-bit and 2-bit cell precisions,  considering the process variation. Comparing to other emerging designs such as 22nm  FeFET \cite{ni2018memory}, 32nm TaOx/HfOx RRAM \cite{hfox},  PCM \cite{kim2019confined}, 
and HZO FeFET \cite{ni2018memory},
as well as the 45nm saturated and subthreshold 1FeFET-1R \cite{Soliman_2020} and 2T-1FeFET \cite{zhou2024low}, the benchmarking results show that our proposed 2FeFET-1T TReCiM array achieves 26.06 TOPS/W energy efficiency at 1-bit precision, which is approximately $2\times$ that of the saturated 1FeFET-1R CiM design and comparable to designs with smaller technology feature sizes. 
With 2-bit cell precision, the proposed TReCiM array achieves  48.03 TOPS/W energy efficiency.
\section{Conclusion}
\label{sec:conclusion}
In this work, 
we first investigate the impact of temperature drift on FeFET based CiM structures. 
To address the potential computation failure caused by the CiM output overlaps due to  temperature drift, we propose TReCiM, a novel subthreshold-FeFET based multibit CiM design that consumes ultra low power  while being resilient to temperature drift. 
The cell and array structures as well as the multibit MAC operations of the proposed design are introduced, and  improved temperature resilience is validated.
Our design shows $3\times$ improvement on temperature resilience compared to the 1FeFET-1R design.
We then 
revise the NeuroSim simulator to  benchmark our proposed design in the context of VGG-8 neural network model running CIFAR-10 dataset, and results suggest that our proposed TReCiM array achieves
an accuracy of $91.31\%$ with 2-bit, and $92.00\%$ with binary storage considering the impacts of process variation. 
Furthermore, our design achieves 48.03 TOPS/W energy efficiency at $128\times 128$ array scale with 2-bit cells. Overall, our proposed design mitigates the temperature impacts, ensures reliable multibit MAC operations and improves the computing efficiency for practical edge neural networks.

\section*{Acknowledgements}
This work was supported in part by  National Key Research and Development Program of China (2022YFB4400300),  NSFC (62104213, 92164203) and SGC Cooperation Project (Grant No. M-0612). 

\small
{

\bibliographystyle{IEEEtran}
\bibliography{bib}
}
\end{document}